\documentstyle[12pt]{article}

\font\twlgot =eufm10 scaled \magstep1
\font\egtgot =eufm8
\font\sevgot =eufm7

\font\twlmsb =msbm10 scaled \magstep1
\font\egtmsb =msbm8
\font\sevmsb =msbm7

\newfam\gotfam

\textfont\gotfam\twlgot
\scriptfont\gotfam\egtgot
\scriptscriptfont\gotfam\sevgot

\newfam\msbfam
\textfont\msbfam\twlmsb
\scriptfont\msbfam\egtmsb
\scriptscriptfont\msbfam\sevmsb
\def\Bbb{\protect\pBbb}
\def\pBbb{\relax\ifmmode\expandafter\Bb\else\typeout{You cann't use
Bbb in text mode}\fi}
\def\Bb #1{{\fam\msbfam\relax#1}}

\def\thebibliography#1{\bigskip\section*{}\bigskip\list
{$^{\arabic{enumi}}$}{\settowidth\labelwidth{#1}\leftmargin\labelwidth
\advance\leftmargin\labelsep
\usecounter{enumi}}
\def\newblock{\hskip .11em plus .33em minus .07em}
\sloppy\clubpenalty4000\widowpenalty4000
\sfcode`\.=1000\relax}

\def\op#1{\mathop{{\it\fam0} #1}\limits}

\newcommand{\im}{{\rm Im\, }}
\newcommand{\re}{{\rm Re\, }}

\newcommand{\lng}{\langle}
\newcommand{\rng}{\rangle}

\newcommand{\intr}{{\rm Int\,}}

\newcommand{\bite}{\begin{itemize}}
\newcommand{\eite}{\end{itemize}}
\newcommand{\benu}{\begin{enumerate}}
\newcommand{\eenu}{\end{enumerate}}
\newcommand{\bde}{\begin{description}}
\newcommand{\ede}{\end{description}}
\newcommand{\bquo}{\begin{quote}}
\newcommand{\equo}{\end{quote}}
\newcommand{\bquot}{\begin{quotation}}
\newcommand{\equot}{\end{quotation}}

\newcommand{\eqref}[1]{(\ref{#1})}
\newcommand{\beq}{\begin{equation}}
\newcommand{\eeq}{\end{equation}}
\newcommand{\ben}{\begin{eqnarray}}
\newcommand{\een}{\end{eqnarray}}
\newcommand{\be}{\begin{eqnarray*}}
\newcommand{\ee}{\end{eqnarray*}}
\newcommand{\bea}{\begin{eqalph}}
\newcommand{\eea}{\end{eqalph}}

\newcommand{\cJ}{{\cal J}}

\newcommand{\cE}{{\cal E}}

\newcommand{\ccG}{{\cal G}}

\newcommand{\bb}{{\bf 1}}

\newcommand{\al}{\alpha}

\newcommand{\dl}{\delta}
\newcommand{\la}{\lambda}

\newcommand{\f}{\phi}
\newcommand{\vf}{\varphi}

\newcommand{\om}{\omega}

\newcommand{\m}{\mu}

\newcommand{\g}{\gamma}
\newcommand{\G}{\Gamma}

\newcommand{\th}{\theta}

\newcommand{\si}{\sigma}

\newcommand{\wt}{\widetilde}
\newcommand{\wh}{\widehat}
\newcommand{\ol}{\overline}

\newcommand{\dr}{\partial}

\newcommand{\ot}{\otimes}
\newcommand{\ap}{\approx}

\let\ssection=\section
\renewcommand{\section}{\setcounter{equation}{0}\ssection}

\newcounter{eqalph}
\newcounter{equationa}
\newcounter{remark}
\newcounter{example}
\newcounter{theorem}
\newcounter{proposition}
\newcounter{lemma}
\newcounter{corollary}
\newcounter{definition}
\newenvironment{eqalph}{\stepcounter{equation}
\setcounter{equationa}{\value{equation}}
\setcounter{equation}{0}

\begin{eqnarray}}{\end{eqnarray}\setcounter{equation}{\value{equationa}}}

\def\theremark{\arabic{remark}}
\def\therexample{\arabic{remark}}

\def\thedefinition{\arabic{definition}}

\newenvironment{rem}{\refstepcounter{remark}\medskip\noindent{\it Remark
\theremark:}}{\medskip}

\newenvironment{theo}{\refstepcounter{definition} \medskip
\noindent{\it Theorem \thedefinition:}}{\medskip}
\newenvironment{prop}{\refstepcounter{definition} \medskip
\noindent{\it Proposition \thedefinition:}}{\medskip}

\newcommand{\mar}[1]{}

\begin{document}
\hbox{}

{\parindent=0pt

{\large\bf Nonequivalent representations of nuclear algebras of canonical
commutation relations. Quantum fields}
\bigskip

{\sc Gennadi 
Sardanashvily\footnote{Electronic mail: sard@grav.phys.msu.su}}
\medskip

{\sl 
Department of Theoretical Physics, Physics Faculty, Moscow State
University, 117234 Moscow, Russia}
\bigskip

Non-Fock representations of the canonical
commutation relations modeled over an infinite-dimensional nuclear space
are constructed in an explicit form. The example of the nuclear space of 
smooth real functions of rapid decrease results in nonequivalent
quantizations of scalar fields.
}

\bigskip

\noindent
{\bf I. INTRODUCTION}
\bigskip

By virtue of the well-known Stone--von Neumann uniqueness theorem,
all irreducible representations of the 
canonical commutation relations (henceforth the CCR)
for finite degrees of freedom are equivalent.
On the contrary, the CCR for infinite degrees of freedom admit infinitely many
nonequivalent irreducible representations (see Ref. [1] for a survey).

Ref. [2] provides the comprehensive description of 
representations of the CCR modeled over
an infinite-dimensional nuclear space $Q$. 
These representations are associated
to translationally quasi-invariant measures on the (topological) dual $Q'$ of
$Q$ and, due to the well-known Bochner theorem, are characterized
by continuous positive-definite functions on $Q$.
In Section IV-V of this work, operators of these representations are
written in an explicit form. For instance, the Fock representation  
is associated to a
certain Gaussian measure on $Q'$. If $Q$ is not a nuclear space, 
the Fock representation need not exist (see Remark \ref{ccr2} below). 
 
Nuclear (non-Banach) involutive algebras are widely studied in
algebraic quantum field theory since the well-known GNS construction
for $C^*$-algebras can be generalized to these algebras too.$^{3-5}$
A Banach space is not nuclear, unless it is finite-dimensional (see
Remark \ref{ccr2} below). A physically relevant example of 
an infinite-dimensional nuclear space is the  
space $RS^4$ of smooth real functions of rapid decrease on
$\Bbb R^4$. It is the real subspace of the space $S(\Bbb R^4)$
of smooth complex functions of rapid decrease on $\Bbb R^4$ 
(see Remark \ref{spr451} below). Its 
(topological) dual is the space
$S'(\Bbb R^4)$ of tempered distributions (generalized functions).$^6$
Of course, elements of $RS^4$ by no means are physical fields, but test
functions. Continuous positive forms $f$ on the
tensor Borchers algebra 
\mar{qm801}\beq
A_{RS^4}=\Bbb R\oplus RS^4\oplus RS^8\oplus\cdots \label{qm801}
\eeq
of $RS^4$ are expressed as
\mar{qm810}\beq
f(\psi_n)=
\int W_n(x_1,\ldots,x_n)\psi_n(x_1,\ldots,x_n)d^4x_1\cdots d^4x_n,
\qquad \psi_n\in RS^{4n}, \label{qm810}
\eeq
into the tempered distributions $W_n\in S'(\Bbb R^{4n})$ whose Fourier 
transform is regarded as the vacuum expectations of the plane wave operators
of quantum scalar fields. 

Of course, quantum fields do not
constitute any CCR algebra, but there is a morphism of $RS^4$ to the CCR
algebra over the nuclear space $RS^3$. It is treated as the
instantaneous CCR algebra of scalar fields. Its Fock representation
provides the familiar vacuum expectations of free quantum scalar
fields on the Minkowski space $\Bbb R^4$, while the non-Fock ones lead
to nonstandard quantizations of these fields (see Section VI). 

In order to characterize interacting quantum fields
created at some instant
and annihilated at another one, one should turn to
the causal forms $f^c$ on the Borchers algebra $A_{RS^4}$ (\ref{qm801}).  
They are given by the functionals
\mar{1260,030}\ben
&&f^c(\psi_n)=
\int W_n^c(x_1,\ldots,x_n)\psi_n(x_1,\ldots,x_n)d^4x_1\cdots d^4x_n,
\qquad \psi_n\in RS^{4n}, \label{1260}\\
&& W^c_n(x_1,\ldots,x_n)=
\op\sum_{(i_1\ldots i_n)}\th(x^0_{i_1}-x^0_{i_2})
\cdots\th(x^0_{i_{n-1}}-x^0_{i_n})W_n(x_1,\ldots,x_n), \label{030}
\een
where $W_n\in S'(\Bbb R^{4n})$ are tempered distributions, $\th$ is the step
function, and the sum runs
through all permutations $(i_1\ldots i_n)$ of the tuple of
numbers $1,\ldots,n$.$^7$
The problem is that the functionals $W^c_n$ (\ref{030}) need not
be tempered distributions and, therefore,
the causal forms $f^c$ (\ref{1260}) are
neither positive nor continuous forms on the Borchers algebra $A_{RS^4}$.

At the same time, the causal forms issue
from the Wick rotation of
Euclidean states of the Borchers algebra $A_{RS^4}$
which describe particles
in the interaction zone (see Section VII). The key point is that,
since the causal forms (\ref{030}) are symmetric,
the Euclidean states of the Borchers algebra $A_{RS^4}$
can be obtained as states of the corresponding commutative tensor
algebra $B_{RS^4}$.$^{8-10}$ They characterize different
representations of the Abelian subgroup of the CCR group modeled over
the nuclear space $SR^4$, and are associated to different positive measures on 
the space of generalized functions $S'(\Bbb R^4)$. From the physical
viewpoint, these states are Euclidean
Green's functions whose Wick rotation gives complete Green's
functions of interacting quantum scalar fields on the Minkowski space.
Some nonperturbative phenomena, e.g., the Higgs vacuum can be
studied in this manner.$^9$

For the sake of simplicity, our consideration here is restricted to
scalar fields. 
In order to describe nonscalar fields on $\Bbb R^4$
with values in a vector space $V$, one can consider the
Borchers algebra of the tensor product space $V\ot SR^4$.$^{10}$
Difficulties arise if nonscalar fields are defined 
on a noncontractible manifold $X$, i.e. they are sections of a vector bundle
$Y\to X$. If this is a trivial bundle,  the space $Y_K(X)$ of
its sections of compact support equipped with the Schwartz topology is a 
nuclear space. In the general case, $Y\to X$ 
is a Whitney summand of
a trivial bundle, and the vector space $Y_K(X)$ of its sections of compact
support is provided with the relative Schwartz topology, which makes
it to a nuclear Schwartz manifold. However, the extension of the
Bochner theorem for this nuclear manifold remains under
question. Note that quantum gauge theory on compact manifolds usually
deals with Sobolev spaces of gauge
potentials.$^{11,12}$ To dispose of the compactness assumption, the
technique of nuclear Schwartz manifolds also has been applied to gauge
theory.$^{13,14}$ However, it meets serious inconsistencies because of
the lack of the inverse function theorem.

\bigskip

\noindent
{\bf II. THE NUCLEAR CCR}
\bigskip

Let us recall the notion of a nuclear space.$^{2,15}$ 
Let a complex vector space $Q$ have a countable set of
nondegenerate
Hermitian forms $\lng.|.\rng_k$, $=1,\ldots$, such that
\be
\lng q|q\rng_1\leq \cdots\leq \lng q|q\rng_k\leq\cdots
\ee
for all $q\in Q$. If $Q$ is complete with respect to the (Hausdorff)
topology defined by the set of norms
\mar{spr445}\beq
\|.\|_k=\lng.|.\rng^{1/2}_k, \qquad k=1,\ldots, \label{spr445}
\eeq
it is called a countably Hilbert space.
The dual $Q'$ of $Q$ is provided with the
weak and strong topologies.

Let $Q_k$ denote the completion of $Q$ with respect to the norm
$\|.\|_k$ (\ref{spr445}). We have the chain of injections
\be
Q_1\supset Q_2\supset \cdots Q_k\supset \cdots 
\ee
together with the homeomorphism $Q=\op\cap_k Q_k$.
Let $T^n_m$, $m\leq n$, be a
prolongation of the map
\be
Q_n\supset Q\ni q\mapsto q\in Q\subset Q_m
\ee
to the continuous map of $Q_n$ onto the dense subset of $Q_m$. A
countably Hilbert space $Q$
is called a nuclear space if, for any $m$,
there exists $n$ such that $T^m_n$ is a nuclear map, i.e.,
\be
T^n_m(q)=\op\sum_i\la_i\lng q|q^i_n\rng_{Q_n} q^i_m,
\ee
where: (i)  $\{q^i_n\}$ and $\{q_m^i\}$ are bases for the
Hilbert spaces $Q_n$ and $Q_m$, respectively, (ii) $\la_i\geq 0$, and
(iii) the series
$\sum \la_i$ converges.

\begin{rem} \label{ccr5} \mar{ccr5}
A nuclear space is perfect,
i.e., every bounded closed set in a
nuclear space is compact. It follows that a Hilbert
space is not nuclear, unless it
is finite-dimensional. Furthermore, a nuclear space is
separable, and
the weak and strong topologies both on this space and its dual coincide.
\end{rem}

Let a nuclear space $Q$ be provided with still another nondegenerate
Hermitian form $\lng.|.\rng$ which is separately continuous.
It follows that there exist numbers $M$ and $m$ such that
\mar{1085}\beq
\lng q|q\rng\leq M\|q\|_m, \qquad \forall q\in Q. \label{1085}
\eeq
Let $\wt Q$ denote the completion of $Q$ with respect to this
form. There are the injections
\mar{spr450}\beq
Q\subset \wt Q\subset Q', \label{spr450}
\eeq
where $Q$ is dense in $\wt Q$, and so is $\wt Q$ in
$Q'$. The triple (\ref{spr450}) is
called the rigged Hilbert space. 

Given a real nuclear space $Q$ together with a nondegenerate
separately continuous Hermitian form $\lng.|.\rng$, let us consider the
group $G(Q)$ 
of triples $g=(q_1,q_2,\la)$ of elements $q_1$, $q_2$
of $Q$ and complex numbers $\la$ of unit modulus which are subject to
multiplications
\mar{qm541}\beq
(q_1,q_2,\la)(q'_1,q'_2,\la')=(q_1+q'_1,q_2+q'_2, \exp[i\lng
q_2,q'_1\rng] \la\la'). \label{qm541}
\eeq
It is a Lie group whose
group space is a nuclear manifold modeled over $Q\oplus Q\oplus \Bbb R$.
Let us denote
\be
T(q)=(q,0,0),  \qquad P(q)=(0,q,0).
\ee
Then the multiplication law (\ref{qm541}) takes the form
\mar{qm543}\ben
&& T(q)T(q')=T(q+q'),\qquad P(q)P(q')=P(q+q'),\nonumber\\
&& P(q)T(q')=\exp[i\lng q|q'\rng]T(q')P(q). \label{qm543}
\een
Written in this form, $G(Q)$ is called the nuclear Weyl CCR group. 

The complexified Lie algebra of the nuclear Lie group $G(Q)$ is the 
Heisenberg
CCR algebra $\ccG(Q)$. It is 
generated by the elements
$\f(q)$, $\pi(q)$, $q\in Q$,
and $I$ which obey the Heisenberg
CCR commutation relations 
\mar{qm540}\ben
&& [\f(q),I]=[\pi(q),I]=0,\nonumber\\
&& [\f(q),\f(q')]=[\pi(q),\pi(q')]=0, \qquad
 [\pi(q),\f(q')]=-i\lng q|q'\rng I. \label{qm540}
\een
There is the exponential map
\be
T(q)=\exp[i\f(q)], \qquad P(q)=\exp[i\pi(q)].
\ee

Due to the relation (\ref{1085}), 
the normed topology on the pre-Hilbert space $Q$ defined by the
Hermitian form $\lng.|.\rng$
is coarser than the 
nuclear space topology. The latter is metric, separable and,
consequently, second-countable. Hence, the pre-Hilbert space
$Q$ is also second-countable and, therefore, admits a countable
orthonormal basis.
Given such a basis $\{q_i\}$ for $Q$,
the Heisenberg CCR (\ref{qm540}) take the form
\be
[\f(q_j),\f(q_k)]=[\pi(q_k),\pi(q_j)]=0, \qquad
[\pi(q_j),\f(q_k)]=-i\dl_{jk}I. 
\ee

\bigskip

\noindent
{\bf III. REPRESENTATIONS OF THE NUCLEAR CCR GROUP}
\bigskip

The CCR group $G(Q)$
contains two nuclear Abelian subgroups $T(Q)$ and $P(Q)$. 
Following the
representation algorithm in Ref. [2], we first construct representations
of the nuclear Abelian group $T(Q)$.  These representations under certain
conditions can be extended to representations of the whole CCR group $G(Q)$. 

One can think of the nuclear Abelian group $T(Q)$ as being the group of
translations in the nuclear space $Q$. 
Its cyclic strongly continuous unitary representation $\pi$
in a Hilbert space
$(E,\lng.|.\rng_E)$ with a (normed) cyclic vector $\theta\in E$ defines  
the complex function 
\be
Z(q)=\lng \pi(T(q))\theta|\theta\rng_E
\ee
on $Q$. This function is proved to be 
continuous and positive-definite, i.e., $Z(0)=1$ and
\be
\op\sum_{i,j} Z(q_i-q_j)\ol c_i c_j\geq 0 
\ee
for any finite set $q_1,\ldots,q_m$ of elements of $Q$ and arbitrary
complex numbers $c_1,\ldots,c_m$. 

In accordance with the well-known  Bochner theorem for nuclear spaces, 
any continuous positive-definite function $Z(q)$
on a nuclear space $Q$ is the Fourier
transform
\mar{qm545}\beq
Z(q)=\int\exp[i\lng q,u\rng]\m(u) \label{qm545}
\eeq
of a positive measure $\m$ of total mass 1 on the dual $Q'$ of
$Q$.$^2$ Then the above mentioned 
representation $\pi$ of $T(Q)$ can be given by the operators
\mar{q2}\beq
T_Z(q)\rho(u)=\exp[i\langle q,u\rangle]\rho(u)  \label{q2}
\eeq
in the Hilbert space $L_C^2(Q',\m)$ of classes of $\m$-equivalent
square integrable complex
functions $\rho(u)$ on $Q'$.
The cyclic vector $\th$ of this representation is the 
$\m$-equivalence class $\th\ap_\m 1$ of the constant function $\rho(u)=1$.
Then we have 
\mar{ccr1}\beq
Z(q)=\lng T_Z(q)\th|\th\rng_\m=\int \exp[i\langle q,u\rangle]\m. \label{ccr1}
\eeq

Conversely, every positive measure $\m$ of total mass 1 on the dual $Q'$
of $Q$ defines the cyclic strongly continuous unitary representation
(\ref{q2}) of 
the group $T(Q)$. By virtue of the above mentioned Bochner theorem, it
follows that every continuous positive-definite 
function $Z(q)$ on $Q$ characterizes a cyclic strongly continuous
unitary representation (\ref{q2}) of  
the nuclear Abelian group $T(Q)$.
We agree to call $Z(q)$ a
generating function of this representation.

It should be emphasized that the representation (\ref{q2})
need not be (topologically) irreducible. For instance, let $\rho(u)$ 
be a function
on $Q'$ such that the set where it vanishes is not a $\m$-null subset of $Q'$.
Then the closure
of the set $T_Z(Q)\rho$ is a $T(Q)$-invariant closed
subspace of $L_C^2(Q',\m)$.

One can show that distinct generating functions $Z(q)$ and $Z'(q)$ determine
equivalent representations $T_Z$ and $T_{Z'}$ (\ref{q2}) of $T(Q)$ in
the Hilbert spaces $L^2_C(Q',\m)$ and $L^2_C(Q',\m')$  iff they are
the Fourier transform of equivalent measures $Q'$.$^2$ Indeed, let
\mar{qm581}\beq
\m'= s^2\m, \label{qm581}
\eeq
where a function $s(u)$ is strictly positive almost everywhere on
$Q'$, and $\m(s^2)=1$. Then the map
\mar{qm580}\beq
L^2_C(Q',\m')\ni\rho(u)\mapsto s(u)\rho(u)\in L^2_C(Q',\m) \label{qm580}
\eeq
provides an isomorphism between the representations $T_{Z'}$ and 
$T_Z$.

The representation $T_Z$ (\ref{q2}) of the nuclear Abelian group
$T(Q)$ in the Hilbert space $L^2_C(Q',\m)$ determined by
the generating function $Z$ (\ref{qm545}) can be extended to
the CCR group $G(Q)$ if the measure $\m$ possesses the following property.

Let $u_q$, $q\in Q$, be an element of $Q'$ given by the condition
\mar{qm546}\beq
\lng q',u_q\rng=\lng q'|q\rng, \qquad \forall q'\in Q. \label{qm546}
\eeq
These elements form the image of the monomorphism
$Q\to Q'$ determined by the Hermitian form $\lng.|.\rng$ on $Q$.
Let the measure $\m$ in (\ref{qm545}) remains equivalent under translations
\be
Q'\ni u\mapsto u+u_q \in Q', \qquad  \forall u_q\in Q\subset Q',
\ee
in $Q'$, i.e.,
\mar{qm547}\beq
\m(u+u_q)=a^2(q,u)\m(u), \qquad \forall u_q\in Q\subset Q', \label{qm547}
\eeq
where a function $a(q,u)$ is square $\m$-integrable and strictly
positive almost everywhere on $Q'$. This function fulfils the relations
\mar{qm555}\beq
a(0,u)=1, \qquad a(q+q',u)=a(q,u)a(q',u+u_q). \label{qm555}
\eeq
A measure on $Q'$ obeying the condition (\ref{qm547}) is called 
translationally quasi-invariant, but
it does not remains equivalent under any translation
in $Q'$, unless $Q$ is finite-dimensional.

Let a generating function $Z$ of a cyclic strongly continuous unitary
representation of the nuclear group $T(Q)$ be the Fourier transform
(\ref{qm545}) of a translationally
quasi-invariant measure $\m$ on $Q'$.  Then one can extend the representation
(\ref{q2}) of this group to the representation of the CCR
group in the Hilbert space $L^2_C(Q',\m)$ by operators 
\mar{qm548}\beq
P_Z(q)\rho(u)=a(q,u)\rho(u+u_q). \label{qm548}
\eeq
Indeed, it is easily 
justified that the
Weyl CCR (\ref{qm543}) hold, while the equalities
\mar{qm550}\ben
&& \|\rho\|_\m=\int |\rho(u)|^2\m(u)=\int |\rho(u+u_q)|^2 \m(u+u_q)=
\label{qm550}\\
&& \qquad \int
a^2(q,u)|\rho(u+u_q)|^2 \m(u)=\|P_Z(q)\rho\|^2_\m, \nonumber
\een
show that the operators (\ref{qm548}) are unitary. 

Let $\m'$ (\ref{qm581}) be a $\m$-equivalent positive measure 
of total mass 1 on $Q'$. The equality
\be
\m'(u+u_q)=s^{-2}(u)a^2(q,u)s^2(u+u_q)\m'(u)
\ee
shows that it is also translationally quasi-invariant. Then the isomorphism
(\ref{qm580}) between representations $T_Z$ and $T_{Z'}$ of the 
nuclear Abelian group
$T(Q)$ is extended to the isomorphism 
\be
P_{Z'}(q)= s^{-1}P_Z(q)s: \rho(u)\mapsto s^{-1}(u)a(q,u)s(u+u_q)\rho(u+u_q)
\ee
of the corresponding representations of the CCR group $G(Q)$.

\bigskip

\noindent
{\bf IV. REPRESENTATIONS OF THE CCR ALGEBRA}
\bigskip

Similarly to the case of a finite-dimensional Lie group,
any strongly continuous
unitary representation $T_Z$ (\ref{q2}), $P_Z$ (\ref{qm548}) of the
nuclear CCR group $G(Q)$ implies a  
representation
of its Lie algebra $\ccG(Q)$ by (unbounded) operators in the same
Hilbert space $L^2_C(Q',\m)$.  This representation reads$^{16,17}$
\mar{qm549,62}\ben
&& I=\bb,\quad \f(q)\rho(u)=\lng q,u\rng\rho(u), \quad
  \pi(q)\rho(u)=-i(\dl_q+\eta(q,u))\rho(u), \label{qm549}\\
&& \dl_q\rho(u)=\op\lim_{\al\to 0}\al^{-1}[\rho(u+\al u_q)-\rho(u)],
\qquad \al\in\Bbb R,\nonumber\\
&& \eta(q,u)=\op\lim_{\al\to 0}\al^{-1}[a(\al q,u)-1].\label{qm562}
\een
One at once derives from the relations (\ref{qm555}) that 
\be
&& \dl_q\dl_{q'}=\dl_{q'}\dl_q, \qquad
\dl_q(\eta(q',u))=\dl_{q'}(\eta(q,u)),\\
&& \dl_q=-\dl_{-q}, \qquad \dl_q(\lng q',u\rng)=\lng q'|q\rng,\\
&& \eta(0,u)=0, \quad \forall u\in Q',\qquad
\dl_q\th=0,  \quad \forall q\in Q.
\ee
With the aid of these relations, it is easily justified that the operators
(\ref{qm549}) fulfil the Heisenberg CCR (\ref{qm540}).
The unitarity condition (\ref{qm550}) implies the conjugation rule
\be
\lng q,u\rng^*=\lng q,u\rng, \qquad \dl_q^*=-\dl_q -2\eta(q,u). 
\ee
Hence, the operators (\ref{qm549}) are Hermitian.

Let us further restrict our consideration to representations with generating
functions $Z(q)$ such that 
\mar{qm586}\beq
\Bbb R\ni t\to Z(tq) \label{qm586}
\eeq
is an analytic function on $\Bbb R$ at $t=0$ for all $q\in Q$.
Then one can show that the function $\lng q|u\rng$ on $Q'$
is square $\m$-integrable for all $q\in Q$ and that, consequently,
the operators
$\f(q)$ (\ref{qm549}) are bounded everywhere in the Hilbert space
$L^2_C(Q',\m)$. Moreover, the mean values of operators $\f(q)$ can be
computed by the formula
\mar{ccr8}\beq
\lng \f(q_1)\cdots\f(q_n)\rng=i^{-n}
\frac{\dr}{\dr\al^1}
\cdots\frac{\dr}{\dr\al^n}Z(\al^iq_i)|_{\al^i=0}=
\int\lng q_1,u\rng\cdots\lng q_n, u
\rng \mu(u). \label{ccr8}
\eeq

The operators $\pi(q)$ (\ref{qm549}) act in the subspace $E_\infty$
of all smooth complex functions in $L^2_C(Q',\m)$ whose derivatives of any
order also belongs to $L^2_C(Q',\m)$. However, $E_\infty$
need not be dense
in the Hilbert space $L^2_C(Q',\m)$, unless $Q$ is finite-dimensional.
The space $E_\infty$ is also the carrier space of a representation of
the enveloping algebra $\ol\ccG(Q)$ of the CCR algebra $\ccG(Q)$.
The representations of $\ccG(Q)$ and $\ol\ccG(Q)$
in $E_\infty$ need not be irreducible. Therefore, let us consider
the subspace $E_\th=\ol\ccG(Q)\th$ of $E_\infty$, where $\th$ is a cyclic
vector for the representation of the CCR group in $L^2_C(Q',\m)$.
Obviously, the representation of the CCR algebra $\ccG(Q)$
in $E_\th$
is (algebraically) irreducible. If $\th'$ is another cyclic vector
in $L^2_C(Q',\m)$, the representations of $\ccG(Q)$
in $E_\th$ and $E_{\th'}$ are equivalent. 

One also introduces creation and
annihilation operators 
\mar{qm552}\beq
a^\pm(q)=\frac{1}{\sqrt 2}[\f(q)\mp i\pi(q)]=\frac{1}{\sqrt 2}[\mp\dl_q
\mp\eta(q,u) + \lng q,u\rng]. \label{qm552}
\eeq
They obey the conjugation rule $(a^\pm(q))^*=a^\mp(q)$
and the commutation relations
\be
[a^-(q), a^+(q')]=\lng q|q'\rng\bb, \qquad
[a^+(q),a^+(q')]=[a^-(q),a^-(q')]=0.  
\ee
The particle number operator
$N$ in the carrier space $E_\th$ is defined by the conditions
\be
[N,a^\pm(q)]=\pm a^\pm(q)  
\ee
up to a summand $\la\bb$. With respect to a countable orthonormal
basis $\{q_k\}$, this operator $N$ is given by the sum
\mar{qm591}\beq
N=\op\sum_k a^+(q_k)a^-(q_k), \label{qm591}
\eeq
but need not be defined everywhere in $E_\th$, unless $Q$ is finite 
dimensional.

\bigskip

\noindent
{\bf V. NON-FOCK REPRESENTATIONS OF THE NUCLEAR CCR}
\bigskip

Gaussian measures exemplifies the physically relevant class of
translationally quasi-invariant measures on the dual $Q'$ of a nuclear
space $Q$. The Fourier transform of a Gaussian measure reads
\mar{spr523}\beq
Z(y)=\exp\left[-\frac12 B(y)\right], \label{spr523}
\eeq
where $B(q)$ is a seminorm on $Q'$ called the covariance form.

\begin{rem} \label{ccr2} \mar{ccr2}
If $Q$ is a Banach space provided with the norm
$\|.\|$, there exists a Gaussian quasi-measure on its dual $Q'$ with the
covariance form $\|.\|$, but it is not a measure unless $Q$ is
finite-dimensional. Let $T$ be a continuous operator in $Q$.
The Gaussian quasi-measure on $Q'$ with the
covariance form $q\mapsto\|Tq\|$ is proved to be a measure iff
$T$ is a Hilbert--Schmidt operator. Let $Q=\Bbb R^n$ be a
finite-dimensional vector space and $B$ a norm on $Q$.
Let its dual $Q'$ be coordinated by $(x_i)$. The Gaussian measure on
$Q'$ with the 
covariance form $B$ is equivalent to the Lebesgue measure $d^nx$ on
$Q'$. It reads 
\be
\m_B= \frac{{\rm det}[B]^{1/2}}{(2\pi)^{n/2}}
\exp\left[-\frac12 (B^{-1})^{ij}x_ix_j\right]d^nx. 
\ee
\end{rem}

Let $\m_K$ denote a Gaussian measure on $Q'$ 
whose Fourier transform is the generating function
\mar{qm563}\beq
Z_K=\exp[-\frac12 B_K(q)] \label{qm563}
\eeq
with the
covariance form
\mar{qm560}\beq
B_K(q)=\lng K^{-1}q|K^{-1}q\rng, \label{qm560}
\eeq
where $K$ is a bounded invertible operator in the Hilbert completion $\wt Q$
of $Q$ with respect to the Hermitian form $\lng.|.\rng$.
The Gaussian measure
$\m_K$ is translationally quasi-invariant, i.e.,
\be
\m_K(u+u_q)=a_K^2(q,u)\m_K(u).
\ee
Using the formula (\ref{ccr8}), one can show that 
\mar{qm561}\beq
a_K(q,u)= \exp[-\frac14 B_K(Sq)-
\frac12\lng Sq,u\rng], \label{qm561} 
\eeq
where $S=KK^*$ is a bounded Hermitian operator in $\wt Q$. 

Let us construct the representation of the CCR algebra $\ccG(Q)$
determined by generating the function $Z_K$ (\ref{qm563}).
Substituting the function (\ref{qm561})
into the formula (\ref{qm562}), we find
\be
\eta(q,u)= -\frac12\lng Sq,u\rng. 
\ee
Hence, the operators $\f(q)$ and $\pi(q)$ (\ref{qm549}) take the form
\mar{qm565}\beq
\f(q)=\lng q,u\rng, \qquad
\pi(q)=-i(\dl_q-\frac12\lng Sq,u\rng). \label{qm565}
\eeq
Accordingly, the creation and annihilation operators (\ref{qm552})
read
\mar{qm566}\beq
a^\pm(q)=\frac{1}{\sqrt 2}[\mp\dl_q \pm \frac12\lng Sq,u\rng + \lng
q,u\rng]. \label{qm566}
\eeq
They act on the subspace $E_\th$, $\th\ap_{\m_K}1$, of the
Hilbert space $L^2_C(Q',\m_K)$,
and are Hermitian with respect to the Hermitian form
$\lng.|.\rng_{\m_K}$ on $L^2_C(Q',\m_K)$.

\begin{rem} \label{ccr10} \mar{ccr10}
If a representation of the CCR is characterized by the Gaussian
generating function (\ref{qm563}), it is convenient for a computation
to express all operator into the operators $\dl_q$ and $\f(q)$, which
obey the commutation 
relation 
\be
[\dl_q,\f(q')]=\lng q'|q\rng.
\ee 
For instance, we have
\be
\pi(q)=-i\dl_q -\frac{i}{2}\f(Sq).
\ee
The mean values $\lng\f(q_1)\cdots\f(q_n)\dl_q\rng$ vanishes, while the
meanvalues $\lng\f(q_1)\cdots\f(q_n)\dl_q\rng$, defined by the
formula (\ref{ccr8}), obey the Wick theorem relations
\mar{ccr20}\beq
\lng\f(q_1)\cdots\f(q_n)\dl_q\rng =\sum \lng\f(q_{i_1})\f(q_{i_2})\rng
\cdots \lng\f(q_{i_{n-1}})\f(q_{i_n})\rng, \label{ccr20}
\eeq
where the sum runs through all partitions of the set $1,\ldots,n$ in
ordered pairs $(i_1<i_2),\ldots(i_{n-1}<i_n)$, and where
\be
\lng\f(q)\f(q')\rng=\lng K^{-1}q|K^{-1}q'\rng.
\ee
\end{rem}

In particular, put $K=\sqrt2\cdot\bb$.
Then the generating function (\ref{qm563}) takes the form
\mar{qm567}\beq
Z_{\rm F}(q)=\exp[-\frac14\lng q|q\rng], \label{qm567}
\eeq
and determines the Fock
representation
of the CCR algebra $\ccG(Q)$. It is given by the operators
\be
&& \f(q)=\lng q,u\rng, \qquad  \pi(q)=-i(\dl_q-\lng q,u\rng), \\
&& a^+(q)=\frac{1}{\sqrt 2}[-\dl_q + 2\lng q,u\rng], \qquad
a^-(q)=\frac{1}{\sqrt 2}\dl_q. 
\ee
Its carrier space is the subspace $E_\th$, $\th\ap_{\m_{\rm F}}1$
  of the Hilbert space
$L^2_C(Q',\m_{\rm F})$, where $\m_{\rm F}$ denotes the Gaussian
measure whose Fourier transform is (\ref{qm567}). We agree to
call it the Fock measure. 

The Fock representation up to an equivalence is characterized by
the existence of a cyclic vector
$\th$ such that
\mar{qm570}\beq
a^-(q)\th=0, \qquad \forall q\in Q. \label{qm570}
\eeq
For the representation in question, this is $\th\ap_{\m_{\rm F}}1$.
An equivalent condition is that the particle number operator $N$ (\ref{qm591})
exists and its spectrum is lower bounded. The corresponding eigenvector
of $N$ in $E_\th$ is $\th$ itself so that $N\th=0$. Therefore,
one often interprets this eigenvector as a vacuum state.

A glance at the expression (\ref{qm566}) shows that the condition
(\ref{qm570}) does not hold, unless $Z_K$ is $Z_{\rm F}$ (\ref{qm567}).
For instance, the particle number operator in the representation
(\ref{qm566}) reads
\be
&& N=\op\sum_j a^+(q_j)a^-(q_j)= \op\sum_j[-\dl_{q_j}\dl_{q_j}
+S^j_k\lng q_k,u\rng\dr_{q_j} + \\
&& \qquad
(\dl_{km}-\frac14 S^j_kS^j_m\lng q_k,u\rng\lng q_m,u\rng -
(\dl_{jj}-\frac12 S^j_j)], 
\ee
where $\{q_k\}$ is the orthonormal basis for the pre-Hilbert space $Q$.
One can show that this operator is
defined everywhere on $E_\th$ and is lower bounded  only if the operator $S$
is a sum of the scalar operator $2\cdot\bb$ and a nuclear operator in 
$\wt Q$, 
in particular, if
\be
{\rm Tr}(\bb -\frac12 S)<\infty.
\ee
This condition is also sufficient
for the measures $\m_K$ and $\m_F$ (and, consequently, the corresponding
representations) to be equivalent.$^2$
For instance, the generating function
\be
Z_c(q)=\exp[-\frac{c^2}{2}\lng q|q\rng], \qquad c^2\neq \frac12,
\ee
determines a non-Fock representation
of the nuclear CCR.

\begin{rem} \label{1097} \mar{1097}
The non-Fock representation (\ref{qm565}) of the
CCR algebra (\ref{qm540}) in the Hilbert
space $L^2_C(Q',\m_K)$ is the Fock representation
\be
&& \f_K(q)=\f(q)=\lng q,u\rng,\\
&& \pi_K(q)=\pi(S^{-1}q) = -i(\dl^K_q -\frac12\lng q,u\rng), \qquad
\dl^K_q =\dl_{S^{-1}q},
\ee
of the CCR algebra $\{\f_K(q),\pi_K(q),I\}$, where
\be
[\f_K(q),\pi_K(q)]=i\lng K^{-1}q|K^{-1}q'\rng I.
\ee
This fact motivates somebody to regard
representations of the
CCR group (\ref{qm543}) as
representation of two mutually commutative
Abelian groups $T(Q)$ and $(P(Q)$ up to phase multipliers.$^{18}$
\end{rem}

Since the Fock measure $\m_{\rm F}$ on $Q'$ remains equivalent only
under translations by vectors
$u_q\in Q\subset Q'$, the measure
\be
\m_\si =\m_{\rm F}(u-\si), \qquad \si\in Q'\setminus Q, 
\ee
on $Q'$ determines a non-Fock representation of the nuclear CCR.
Indeed, this measure is translationally quasi-invariant:
\be
\m_\si(u+u_q)=a^2_\si(q,u)\m_\si(u),\qquad 
a_\si(q,u)=a_{\rm F}(q,u-\si), 
\ee
and its Fourier transform
\be
Z_\si(q)=\exp[i\lng q,\si\rng]Z_{\rm F}(q) 
\ee
is a positive-definite continuous function on $Q$.
Then the corresponding
representation of the CCR algebra is given by operators
\mar{qm597}\beq
a^+(q)=\frac{1}{\sqrt 2}(-\dl_q + 2\lng q,u\rng -\lng q,\si\rng),\qquad
a^-(q)=\frac{1}{\sqrt 2}(\dl_q + \lng q,\si\rng). \label{qm597}
\eeq
In comparison with all the above representations,
these operators possess nonvanishing vacuum mean values
\be
\lng a^\pm(q)\th|\th\rng_{\m_{\rm F}}=\mp\lng q,\si\rng.
\ee
If $\si\in Q\subset Q'$, the representation (\ref{qm597})
becomes equivalent to
the Fock representation (\ref{qm566}) due to the morphism
\be
\rho(u)\mapsto \exp[-\lng q',u\rng]\rho(u+u_{q'}).
\ee

\bigskip

\noindent
{\bf VI. FREE QUANTUM FIELDS}
\bigskip

In this Section, representations of the nuclear CCR
are utilized
in order to describe free quantum fields.
In the framework of algebraic quantum field theory, quantum fields
are characterized by a unital involutive topological algebra $A$ and a
(continuous positive) state $f$ of $A$.
The key point is that a quantum field algebra is never normed. 

With reference to the field-particle dualism, realistic quantum field
models are described by tensor algebras, as a rule.
Let $Q$ be a real (locally convex) topological vector space, endowed with an
involution operation $q\mapsto q^*$, $q\in Q$. Let 
us consider
the tensor algebra 
\mar{spr620}\beq
A_Q=\Bbb R\oplus Q \oplus Q^2\oplus\cdots, \qquad Q^n= \op\ot^nQ, \label{spr620}
\eeq
of $Q$. It is a $*$-algebra with
respect to the involution
\be
(q^1\cdots q^n)^*=(q^n)^*\cdots (q^1)^*.
\ee
The direct sum topology makes $A_Q$ to a topological
involutive algebra. A state $f$ of this algebra is given by a tuple $\{f_n\}$
of continuous forms on the tensor products $Q^n$.
Its value $f(q^1\cdots q^n)$ are interpreted as the vacuum expectation of
the system of fields $q^1,\ldots,q^n$. Further, we choose by $Q$  the
real subspace $SR^4$ of the nuclear space of smooth complex functions of rapid
decrease on $\Bbb R^4$.

\begin{rem} \label{spr451} \mar{spr451}
By functions of rapid decrease on an Euclidean space $\Bbb R^n$
are called  complex smooth functions $\psi(x)$ such that
the quantities
\mar{spr453}\beq
\|\psi\|_{k,m}=\op\max_{|\al|\leq k} \op\sup_x(1+x^2)^m|D^\al \psi(x)|
\label{spr453}
\eeq
are finite for all $k,m\in \Bbb N$. Here, we follow 
the standard notation
\be
D^\al=\frac{\dr^{|\al|}}{\dr^{\al_1} x^1\cdots\dr^{\al_n}x^n}, \qquad
|\al|=\al_1+\cdots +\al_n, 
\ee
for an $n$-tuple  of natural numbers $\al=(\al_1,\ldots,\al_n)$.
The functions of rapid decrease constitute the nuclear space
$S(\Bbb R^n)$ with respect to the
topology determined by the seminorms (\ref{spr453}). Its dual
is the space $S'(\Bbb R^n)$ of tempered distributions.$^{2,6,15}$ The
corresponding contraction form is written as
\be
\lng \psi,h\rng=\op\int \psi(x) h(x) d^nx, \qquad \psi\in S(\Bbb R^n),
\qquad h\in S'(\Bbb R^n).
\ee
The  space
$S(\Bbb R^n)$ is provided with the nondegenerate
separately continuous Hermitian form
\be
\lng \psi|h\rng=\int \psi(x)\ol{h(x)}d^nx. 
\ee
The completion of $S(\Bbb R^n)$
with respect to this form is the space $L^2_C(\Bbb R^n)$ of
square integrable complex functions on $\Bbb R^n$.
We have the rigged Hilbert space
\be
S(\Bbb R^n)\subset L^2_C(\Bbb R^n) \subset S'(\Bbb R^n).
\ee
Let $\Bbb R_n$ denote the dual of $\Bbb R^n$  coordinated by
$(p_\la)$. The Fourier transform
\mar{spr460,1}\ben
&& \psi^F(p)=\int \psi(x)e^{ipx}d^nx, \qquad px=p_\la x^\la,
\label{spr460}\\
&& \psi(x)=\int \psi^F(p)e^{-ipx}d_np, \qquad d_np=(2\pi)^{-n}d^np,
\label{spr461}
\een
defines an isomorphism between the spaces $S(\Bbb R^n)$ and
$S(\Bbb R_n)$.
The Fourier transform of tempered distributions is defined by the
condition
\be
\int h(x)\psi(x)d^nx=\int h^F(p)\psi^F(-p)d_np,
\ee
and is written in the form
(\ref{spr460}) -- (\ref{spr461}).
It provides an isomorphism between the spaces of tempered distributions
$S'(\Bbb R^n)$ and $S'(\Bbb R_n)$.
\end{rem}

The tensor algebra $A_{RS^4}$ of the nuclear space $RS^4$ is called the
Borchers algebra.$^{3,4}$ Since the
subset $\op\ot^nS(\Bbb R^k)$ is dense in
$S(\Bbb R^{kn})$, we henceforth identify $A_{RS^4}$ with the algebra
(\ref{qm801}). 
Then any state $f$ of $A_{RS^4}$ is represented 
by a collection
of tempered distributions $\{W_n\in S'(\Bbb R^{4n})\}$ by the formula
(\ref{qm810}).
Let us focus on the states of the Borchers algebra $A_{RS^4}$ which 
describe free real
scalar fields of mass $m$.

Let us provide the nuclear space $RS^4$ with the positive complex bilinear form
\mar{qm802}\ben
&& (\psi|\psi')=\frac{2}{i}\int \psi(x)D^-(x-y)\psi'(y)d^4xd^4y=\int
\psi^F(-\om,-\op p^\to)\psi'^F(\om,\op p^\to)\frac{d_3p}{\om}, 
\label{qm802}\\
&& D^-(x)=i(2\pi)^{-3}\int \exp[-ipx]\th(p_0)\dl(p^2-m^2)d^4p, \nonumber
\een
where $D^-(x)$ is the negative frequency part of the Pauli--Jordan
function, $p^2$ is the Minkowski square, and
\be
\om=({\op p^\to}^2 +m^2)^{1/2}.
\ee
Since the function $\psi(x)$ is real, its Fourier transform satisfies the
equality $\psi^F(p)=\ol\psi^F(-p)$.

The bilinear form (\ref{qm802}) is degenerate
because the Pauli--Jordan function $D^-(x)$ obeys the mass shell equation
\be
(\Box +m^2)D^-(x)=0.
\ee
It takes nonzero values only at elements $\psi^F\in RS_4$
which are not zero on the mass shell $p^2=m^2$. Therefore, let us consider the
quotient space $\g:RS^4\to RS^4/J$,
where $J=\{\psi\in RS^4\, :\, (\psi|\psi)=0\}$
is the kernel of the square form (\ref{qm802}). The map $\g$ assigns to each
element $\psi\in RS^4$ with the Fourier transform $\psi^F(p_0,\op
p^\to)\in RS_4$
the couple of functions $(\psi^F(\om,\op p^\to),\psi^F(-\om,\op p^\to))$. Let
us equip the factor space $RS^4/J$ with the real bilinear form
\mar{qm803}\ben
&& (\g\psi|\g\psi')_L={\rm Re}(\psi|\psi')= \label{qm803}\\
&& \qquad \frac12\int [\psi^F(-\om,-\op p^\to)\psi'^F(\om,\op p^\to)
+\psi^F(\om,-\op p^\to) \psi'^F(-\om,\op p^\to)]\frac{d_3\op
p^\to}{\om}. \nonumber
\een
Then it is decomposed into the direct sum
$RS^4/\cJ=L^+\oplus L^-$
of the subspaces
\be
L^\pm=\{\psi^F_\pm(\om,\op p^\to)=\frac12(\psi^F(\om,\op p^\to)
\pm\psi^F(-\om,\op p^\to))\},
\ee
which are mutually orthogonal with respect to the bilinear form
(\ref{qm803}).

There exist continuous isometric morphisms
\be
&& \g_+:\psi^F_+(\om,\op p^\to) \mapsto q^F(\op p^\to)=\om^{-1/2}\psi^F_+
(\om,\op p^\to),\\
&& \g_-:\psi^F_-(\om,\op p^\to) \mapsto q^F(\op p^\to)=-i\om^{-1/2}\psi^F_-
(\om,\op p^\to)
\ee
of spaces $L^+$ and $L^-$ to the nuclear space $RS^3$ endowed with
the nondegenerate separately continuous Hermitian form
\mar{qm807}\beq
\lng q|q'\rng=\int q^F(-\op p^\to)q'^F(\op p^\to)d_3p. \label{qm807}
\eeq
It should be emphasized that
the images $\g_+(L^+)$ and $\g_-(L^-)$ in $RS^3$ are not orthogonal
with respect to the scalar form (\ref{qm807}). Combining $\g$ and $\g_\pm$,
we obtain the continuous morphisms $\tau_\pm: RS^4\to RS^3$ given by
the expressions
\be
&& \tau_+(\psi)=\g_+(\g\psi)_+=\frac{1}{2\om^{1/2}}\int[\psi^F(\om,\op
p^\to) + \psi^F(-\om,\op p^\to)]\exp[-i\op p^\to\op x^\to]d_3p,\\
&& \tau_-(\psi)=\g_-(\g\psi)_-=\frac{1}{2i\om^{1/2}}\int[\psi^F(\om,\op
p^\to) - \psi^F(-\om,\op p^\to)]\exp[-i\op p^\to\op x^\to]d_3p.
\ee

Now let us consider the CCR algebra
\mar{1110}\beq
\ccG(RS^3)=\{(\f(q),\pi(q),I),\,q\in RS^3\} \label{1110}
\eeq
modeled over the nuclear space $RS^3$, which is equipped with the
Hermitian form (\ref{qm807}). 
Using the morphisms $\tau_\pm$, let us define the map
\mar{1111}\beq
RS^4\ni \psi \mapsto \f(\tau_+(\psi)) -\pi(\tau_-(\psi))\in \ccG(RS^3).
\label{1111} 
\eeq
With this map, one can think of (\ref{1110})
as being
the algebra of the instantaneous CCR of scalar fields on the Minkowski
space $\Bbb R^4$.
Owing to the map (\ref{1111}), any representation of the nuclear
CCR algebra $\ccG(RS^3)$ determined by a translationally 
quasi-invariant measure
$\m$ on $S'(\Bbb R^3)$ induces a state
\mar{qm805}\beq
f(\psi^1\cdots\psi^n)=\lng\f(\tau_+(\psi^1)) +\pi(\tau_-(\psi^1))]
\cdots [\f(\tau_+(\psi^n)) +\pi(\tau_-(\psi^n))]\rng \label{qm805}
\eeq
on the Borchers algebra $A_{RS^4}$ of scalar fields. Furthermore,
one can justify that the corresponding distributions $W_n$ fulfil
the mass shell equation and that the following commutation relation holds:
\be
W_2(x,y) -W_2(y,x)=-iD(x-y),  
\ee
where
\be
D(x)=i(2\pi)^{-3}\int \exp[-ipx](\th(p_0)-\th(-p_0))\dl(p^2-m^2)d^4p,
\ee
is the Pauli--Jordan commutation function. 
Thus, the states (\ref{qm805})
describe real scalar fields of mass $m$.

For instance, let us take the Fock representation (\ref{qm565})
of the CCR algebra $\ccG(RS^3)$. Using the formulae in Remark \ref{ccr10}
where the form $\lng q|q'\rng$ is given by the expression (\ref{qm807}), 
one observes that the states $f_{\rm F}$ (\ref{qm805})
satisfy the Wick theorem relations
\mar{1263}\beq
f_{\rm F}(\psi^1\cdots\psi^n)=\sum f_2(\psi^{i_1}\psi^{i_2})\cdots
f_2(\psi^{i_{n-1}}\psi^{i_n}), \label{1263}
\eeq
while the state
$f_2$ is given by the Wightman function
\mar{ccr21}\beq
W_2(x,y)=\frac{1}{i}D^-(x-y). \label{ccr21}
\eeq
Thus, the state $f_{\rm F}$ describe standard quantum free scalar fields 
of mass $m$.

Similarly, one can obtain states of the Borchers algebra $A_{RS^4}$
generated by non-Fock representations of
the instantaneous CCR algebra $\ccG(RS^3)$, e.g., if $K^{-1}=c\bb\neq
2^{-1/2}\bb$. These states fail to be given by Wightman functions. In
particular, they are not covariant under time translations.

\bigskip

\noindent
{\bf VII. EUCLIDEAN SCALAR FIELDS}
\bigskip

As was mentioned above, the causal forms (\ref{1260}) on the Borchers
algebra $A_{RS^4}$ are neither positive nor continuous. At the same
time, they issue from the Wick rotation of Euclidean states of the
commutative tensor algebra $B_{RS^4}$. These states play the role of Green's
functions in Euclidean quantum field theory. It should be emphasized
that they
do not coincide with the
Schwinger functions in axiomatic quantum field theory
whose Minkowski partners are the Wightman functions, but not causal
forms (see Section VIII).

Let $Q$ be a real nuclear space as above and $A_Q$
its tensor algebra (\ref{spr620}).
We abbreviate with $B_Q$ the complexified quotient of $A_Q$
with respect to the ideal generated by
the elements $q\ot q'-q'\ot q$ for all $q,q'\in Q$. It is the
commutative tensor algebra of $Q$. Provided with the direct sum topology,
$B_Q$ becomes a topological
involutive algebra. It coincides with
the enveloping algebra of the Lie algebra
of the additive Lie group $T(Q)$ of translations in
$Q$. Therefore, we can obtain the states of the algebra
$B_Q$ by constructing cyclic strongly continuous unitary
representations of the nuclear Abelian group $T(Q)$.
As was stated in Section III,
such a representation is
characterized by a positive-definite continuous generating
function $Z$ on $Q$
which is the Fourier
transform (\ref{qm545}) 
of a bounded positive measure $\mu$ of total mass 1
on the (topological) dual $Q'$ of $Q$.
The corresponding cyclic strongly continuous unitary representation of
the nuclear Abelian group $T(Q)$ is given by the operators
(\ref{q2}) in the Hilbert space $L^2_C(Q',\mu)$ of 
square $\m$-integrable
complex functions $\rho(u)$ on $Q$.
If the function (\ref{qm586}) 
is analytic at $\al=0$ for all $\f\in\Phi$, a state $F(q_1\cdots q_n)$
of $B_Q$ is given by the expression (\ref{ccr8}).

A glance at this expression shows that, in applications
to quantum field theory where $Q=RS^4$, the generating function 
$Z$ plays the role
of a generating functional 
represented by the functional integral
(\ref{qm545}), while the values (\ref{ccr8}) of the state $F$
are vacuum expectations of Euclidean fields.

For instance, let $\m$ be a Gaussian
measure on $Q'$ whose Fourier transform
reads
\be
Z(\vf)=\exp[-\frac12 M(\vf)], 
\ee
where the covariance form $M(\f_1,\f_2)$ is a nondegenerate
separately continuous Hermitian form on $RS^4$.
This generating function defines a Gaussian state
$F$ of the algebra $B_{RS^4}$ such that
\be
F_1(\phi)=0, \qquad F_2(\f_1\f_2)=M(\f_1,\f_2),
\ee
while $F_{n>2}$ obey the Wick relations (\ref{ccr20}).
Furthermore, a covariance form $M$ on $RS^4$
is uniquely determined as
\mar{1270}\beq
M(\f_1,\f_2)=\int W_2(x_1,x_2)\f_1(x_1)\f_2(x_2) \label{1270}
d^nx_1d^nx_2.
\eeq
by a tempered distribution $W_2\in S'(\Bbb R^8)$.

In particular, let 
a tempered distribution $M(\f,\f')$ in the expression (\ref{1270}) be 
Green's function
of some positive elliptic differential operator $\cE$, i.e.,
\be
\cE_{y_1}W_2(y_1,y_2)=\dl(y_1-y_2),
\ee
where $\dl$ is Dirac's $\dl$-function. Then the distribution $W_2$ reads
\mar{1272}\beq
W_2(y_1,y_2)=w(y_1-y_2), \label{1272}
\eeq
and we obtain the form
\be
&& F_2(\f_1\f_2)=M(\f_1,\f_2)=\int w(y_1-y_2)\f_1(y_1)\f_2(y_2) 
d^4y_1 d^4y_2=\\
&& \qquad \int w(y)\f_1(y_1)\f_2(y_1-y)d^4y d^4y_1=\int w(y)\vf(y)d^4y=
\int w^F(q)\vf^F(-q) d_4q, \\
&& y=y_1-y_2, \qquad \vf(y)=\int \f_1(y_1)\f_2(y_1-y)d^4y_1.
\ee
For instance, if
\be
\cE_{y_1} =-\Delta_{y_1}+m^2,
\ee
where $\Delta$ is the Laplacian, then
\mar{1271}\beq
w(y_1-y_2)=\int\frac{\exp(-iq(y_1-y_2))}{q^2+m^2}, \label{1271}
\eeq
where $q^2$ is the Euclidean square, is the propagator
of a massive Euclidean scalar field.

Let the Fourier transform $w^F$ of the distribution $w$ (\ref{1272})
satisfy the condition 
(\ref{036}) below. Then its Wick rotation (\ref{040}) is the functional
\be
\wh w(x)=\th(x)\op\int_{\ol Q_+}w^F(q)\exp(-qx)dq +
\th(-x)\op\int_{\ol Q_-}w^F(q)\exp(-qx)dq
\ee
on scalar fields on the Minkowski space. For instance, let $w(y)$ be the
Euclidean propagator
(\ref{1271}) of a massive scalar field. Then due to the analyticity of
\be
w^F(q)=(q^2+m^2)^{-1}
\ee
on the domain $\im q\cdot \re q>0$, one can show that
$\wh w(x)=-iD^c(x)$ 
where $D^c(x)$ is familiar causal Green's function.

\bigskip

\noindent
{\bf VIII. APPENDIX. THE WICK ROTATION}
\bigskip

Let us describe the above mentioned Wick rotation of Euclidean states
in the previous Section, and compare it with the transition between
Wightman and Schwinger functions.

We start from the basic formulae of the Fourier--Laplace transform.$^6$
It is defined on Schwartz distributions, but
we focus on the tempered ones.

Throughout, $\Bbb R^n_+$ and $\ol\Bbb R^n_+$ denote the subset of points
of $\Bbb R^n$ with strictly positive Cartesian coordinates
and its closure, respectively. 
Let $f\in S'(\Bbb R^n)$ be a tempered distribution and $\G(f)$ the 
convex subset
of points $q\in \Bbb R_n$ such that
\mar{1273}\beq
e^{-qx}f(x)\in S'(\Bbb R^n). \label{1273}
\eeq
In particular, $0\in\G(f)$. Let $\intr\G(f)$ and $\dr\G(f)$ denote
the interior and the boundary of $\G(f)$, respectively.

The Fourier--Laplace (henceforth FL) transform 
of a tempered distribution $f\in S'(\Bbb R^n)$ is said to be the tempered
distribution
\mar{7.2}\beq
f^{FL}(p+iq)=(e^{-qx}f(x))^F(p)=\int f(x)e^{i(p+iq)x} d^nx \in S'(\Bbb R_n),
\label{7.2}
\eeq
which is the Fourier transform of the distribution (\ref{1273}) depending on
$q$ as parameters. One can also think of the FL 
transform (\ref{7.2})
as being the Fourier transform with respect to the complex arguments $k=p+iq$.

If $\intr\G(f)\neq\emptyset$, the FL transform 
$f^{FL}(k)$ is a holomorphic
function of complex arguments $k=p+iq$ on the open tube
$\Bbb R_n +i\intr\G(f)\subset \Bbb C_n$
over $\intr\G(f)$. Moreover, for any compact subset $Q\subset \intr\G(f)$,
there exist strictly positive numbers $A$ and $m$, depending of $Q$ and $f$,
such that
\mar{7.3}\beq
|f^{FL}(p+iq)|\leq A(1+|p|)^m, \qquad p\in \Bbb R_n, \qquad q\in Q. \label{7.3}
\eeq
The evaluation (\ref{7.3}) is 
equivalent to the
fact that the function $h(p+iq)$ defines a family of tempered distributions
$h_q(p)\in S'(\Bbb R_n)$ of the variables $p$ depending continuously on
parameters $q\in S$.

Let us notice that, if $0\in\intr\G(f)$, then
\be
f^{FL}(p+i0)=\op\lim_{q\to 0}f^{FL}(p+iq)
\ee
coincides with the Fourier transform $f^F(p)$ of $f$. The case of
$0\not\in\intr\G(f)$ is more intricate. Let $S$ be a convex domain in 
$\Bbb R^n$
such that $0\in\dr S$, and let $h(p+iq)$ be a holomorphic function on the tube
$T^S$ which defines a family of tempered distributions $h_q(p)\in 
S'(\Bbb R_n)$,
depending on parameters $q$. One says that $h(p+iq)$ has a  
generalized boundary
value $h_0(p)\in S'(\Bbb R_n)$ if, for any
frustum $K^r\subset S\cup \{0\}$ of the cone $K\subset \Bbb R_n$ (i.e.,
$K^r=\{q\in K\,:\,|q|\leq r\}$), one has
\be
h_0(\psi(p))=\op\lim_{|q|\to 0,\,q\in K^r\setminus\{0\}} h_q(\psi(p)) 
\ee
for all functions $\psi\in S(\Bbb R_n)$ of rapid decrease. Then the following
assertion holds.

\begin{prop}
Let $f\in S'(\Bbb R^n)$, $\intr\G(f)\neq \emptyset$ and $0\not\in\intr\G(f)$.
Then a generalized boundary value of the FL transform $f^{FL}(k)$
in $S'(\Bbb R_n)$ exists and coincides with the Fourier transform $f^F(p)$ of
the distribution $f$.
\end{prop}

Let us apply this result to the following important case.
The support of a tempered distribution
 $f$ is defined as the complement
of the maximal open subset $U$ where $f$ vanishes, i.e., $f(\psi)=0$ for all
$\psi\in S(\Bbb R^n)$ of support in $U$. Let $f\in S'(\Bbb R^n)$ is 
of support in
$\ol\Bbb R_+^n$. Then $\ol\Bbb R_{n+}\subset \G(f)$, and the 
FL transform
$f^{FL}$ is a holomorphic function on the tube over $\Bbb R_{n+}$, 
while its generalized
boundary value in $S'(\Bbb R_n)$ is given by the equality
\be
h_0(\psi(p))=\op\lim_{|q|\to 0,\,q\in\Bbb R_{n+}} 
f^{FL}_q(\psi(p))=f^F(\psi(p)), \qquad \forall
\psi\in S(\Bbb R_n).
\ee
Conversely, one can restore a tempered distribution $f$ of
support in $\ol\Bbb R_+^n$ from its FL transform
$h(k)=f^{FL}(k)$ even if this function is known only on
$i\Bbb R_{n+}$. Indeed, the formulae
\mar{7.4,5}\ben
&& \wt h(\phi)=\op\int_{\Bbb R_{n+}}h(iq)\f(q)d_nq= \op\int_{\Bbb R_{n+}}
d_nq \op\int_{\ol\Bbb R^n_+} e^{-qx}f(x)\f(q)d^nx= \label{7.4}\\
&& \qquad \op\int_{\ol\Bbb R^n_+} f(x)\wh\f(x)d^nx, \qquad \f\in S(\Bbb
R_{n+}),\nonumber\\
&& \wh\f(x)=\op\int_{\Bbb R_{n+}} e^{-qx}\f(q)d_nq, \qquad
x\in\ol\Bbb R^n_+, \qquad \wh\f\in S(\ol\Bbb R^n_+), \label{7.5}
\een
define a linear continuous functional $\wt h(q)=h(iq)$ on the
space $S(\Bbb R_{n+})$. It is called the Laplace transform
$f^L(q)=f^{FL}(iq)$ of a tempered distribution $f$.

\begin{rem}
Let us illustrate the restoration of a tempered distribution from the
functional (\ref{7.4}) in the case of $n=1$. Let $f\in S'(\ol\Bbb R_+)$.
Its FL transform reads
\mar{1275}\beq
h(p+iq)=\op\int^\infty_0 e^{i(p+iq)x}f(x)dx, \qquad q>0. \label{1275}
\eeq
Since $h(p+iq)$ (\ref{1275}) has the generalized boundary value
$h(p+i0)$, the $f$ is reconstructed from $\wh h(z)=h(iz)$ by the formulae
\mar{7.6}\beq
f(x)=\frac{1}{2\pi i}\op\int^{0+i\infty}_{0-i\infty}e^{zx}\wt h(z)dz,
\qquad \wt h(0-ip)=f^F(p). \label{7.6}
\eeq
where
\be
\wt h(q)=h(iq)=\op\int^\infty_0e^{-qx}f(x)dx, \qquad q>0,
\ee
is the Laplace transform $f^L$ of $f$.
\end{rem}

The image of the space $S(\Bbb R_{n+})$ with respect to the mapping
$\f(q)\mapsto \wh \f(x)$ (\ref{7.3}) is dense in $S(\ol\Bbb R^n_+)$. 
Then the family
of seminorms $\|\f\|'_{k,m}=\|\wh\f\|_{k,m}$, where $\|.\|_{k,m}$ are seminorms
(\ref{spr453}) on $S(\Bbb R^n)$, determines new coarsen 
topology on $S(\Bbb R_{n+})$ such that the functional (\ref{7.4}) 
remains continuous
with respect to this topology. Then the following is proved.$^6$

\begin{theo} \label{1276} \mar{1276}
The mappings (\ref{7.4}) and (\ref{7.5})
provide one-to-one correspondence between the Laplace transforms 
$f^L(q)=f^{FL}(iq)$
of tempered distributions $f\in S'(\ol\Bbb R^n_+)$ and the elements 
of $S'(\Bbb R_{n+})$
which are continuous with respect to the coarsen topology on $S(\Bbb R_{n+})$.
\end{theo}

This Theorem enables one both to establish the relation between 
Wightman and
Schwinger functions and to define the above mentioned Wick rotation of
Euclidean states

We here address Wightman and Schwinger functions in axiomatic field theory
in order to show the difference between Schwinger functions and
the above states of Euclidean fields.

Recall that Wightman functions are defined as
tempered distributions $W_n\subset S'(\Bbb R^{4n})$
on the Minkowski space which obey the Garding--Wightman axioms
of axiomatic field theory.$^{6,19,20}$
Let us mention
the Poincar\'e covariance axiom, the spectrum condition and the 
locality condition. 

Due to the translation covariance of Wightman functions $W_n$,
there exist tempered distributions
$w_n\in S'(\Bbb R^{4n-4})$ such that
\mar{1278}\beq
W_n(x_1,\ldots,x_n)= w_n(x_1-x_2,\ldots,x_{n-1}-x_n). \label{1278}
\eeq
The spectrum condition implies that the Fourier transform $w^F_n$ of
the distributions $w_n$ (\ref{1278}) are of
support in
    the closed forward light cone $\ol V_+$ in the momentum Minkowski
space $\Bbb R_4$. It follows that
the Wightman function $w_n$ is a generalized boundary value in 
$S'(\Bbb R^{4n-4})$
of the function $(w^F_n)^{FL}$, which is the FL transform
of the function $w^F_n$ with respect to variables $p^i_0$ and which is
holomorphic on the tube $(\Bbb R^4+iV_-)^{n-1}\subset \Bbb C^{4n}$. 
Accordingly,
$W_n(x_1,\ldots,x_n)$ is a generalized boundary value in $S'(\Bbb R^{4n})$
of a function $W_n(z_1,\ldots,z_n)$, holomorphic on the tube
\be
\{z_i\,:\, \im(z_{i+1}-z_i)\in V_-, \,\, \re z_i\in\Bbb R^4\}.
\ee
In accordance with the Lorentz covariance, the Wightman functions admit
an analytic continuation
onto a wider domain in $\Bbb C^{4n}$, called the extended forward tube.
Furthermore, the locality condition implies
that they are symmetric on this domain.

 From now on, let us denote by $X$ the space $\Bbb R^4$ associated to the
real subspace of $\Bbb C^4$ and by $Y$ the space $\Bbb R^4$, coordinated by
$(y^0,y^{1,2,3})$ and associated to the subspace $\wt Y$ of $\Bbb C^4$
whose points possess the coordinates $(iy^0,y^{1,2,3})$. If $X$ is 
the Minkowski
space, then one can think of $Y$ as
being its Euclidean partner. 

Let us consider the subset $\wt Y^n_{\neq}\subset \wt Y\subset \Bbb 
C^{4n}$ which consists
of the points $(z_1,\ldots,z_n)$ such that $z_i\neq z_j$. It belongs the
domain of analiticity of the Wightman function $W_n(z_1,\ldots,z_n)$, 
whose restriction
to $\wt Y^n_{\neq}$ defines the symmetric function
\be
S_n(y_1,\ldots,y_n)=W_n(z_1,\ldots,z_n), \qquad z_i=(iy^0_i,y^{1,2,3}_i),
\ee
on $Y^n_{\neq}$.  It is called the Schwinger function. 
On the domain $Y^n_<$ of points $(y_1,\ldots,y_n)$ such that
$0<y_1^0<\cdots< y_n^0$, the Schwinger function takes the form
\mar{1279}\beq
S_n(y_1,\ldots,y_n)=s_n(y_1-y_2,\ldots,y_{n-1}-y_n), \label{1279}
\eeq
where $s_n$ is an element of the space $S'(Y_-^{n-1})$ which is continuous
with respect to the coarsen topology on
$S(Y_-^{n-1})$. Consequently, in accordance with Theorem \ref{1276} 
and by virtue of the formula
(\ref{7.4}), the Schwinger function $s_n$ (\ref{1279}) can be represented
as
\mar{7.8}\ben
&& s_n(y_1-y_2,\ldots,y_{n-1}-y_n)= \label{7.8}\\
&& \qquad \int\exp[p^j_0(y^0_j-y^0_{j+1})
-i\op\sum^3_{k=1}p^j_k (y^k_j- y^k_{j+1})]w_n^F(p^1,\ldots,p^n)
d_4p^1\cdots d_4p^{n-1}, \nonumber
\een
where $w^F_n\in S'(\ol\Bbb R_{n+})$ is the Fourier transform
of the Wightman function $w_n$, seen as an element of $S'(\Bbb R_{n+})$
of support in the subset $p^i_0\geq 0$.
The formula (\ref{7.8}) enables one to
restore the Wightman functions on the
Minkowski from the Schwinger functions on the Euclidean space.$^{19,20}$

For instance, let us consider Wightman and Schwinger functions $w_2$ and $s_2$
for a massive scalar field. The Schwinger one reads
\be
s_2(y_1-y_2)=\int\frac{e^{-ip(y_1-y_2)}}{q^2+m^2}d_4q,
\ee
where $q^2$ is the Euclidean square and $(y_1^0-y_2^0)<0$.
It coincides with the distribution $w(y_1-y_2)$ (\ref{1271}) 
restricted to the domain $(y^0_1-y^0_2)<0$.
Let us omit a computation under the sign of spatial integrals,
and let us introduce the notation $y^0=y_1^0-y_2^0$ and
\be
M=({\op q^\to}^2 +m^2)^{1/2}.
\ee
Then bearing in mind the formula (\ref{7.6}), we obtain
\be
&& s_2(y^0)=\frac{1}{2\pi}\op\int^\infty_{-\infty}
\frac{e^{-iq_0y^0}}{p^2_0+
M^2}dp_0=\frac{1}{2M}e^{My^0}, \qquad y^0<0,\\
&& w_2(p_0)=\frac{1}{2\pi i}\op\int^{0+i\infty}_{0-i\infty}
e^{-p_0y^0}s_2(y^0)dy^0 =\frac{1}{2\pi i}\op\int^{0+i\infty}_{0-i\infty}
\frac{1}{2M} e^{y^0(M-p_0)}dy^0=\\
&& \quad \frac{1}{2\pi}\op\int^\infty_{-\infty}
\frac{1}{2M} e^{ix(M-p_0)}dx= \frac{1}{2M}\dl(p_0-M)=\dl(p_0^2-M^2),
\quad p_0<0, \quad x=-iy^0.
\ee
It follows that $w_2(p)$ is the Fourier transform of the Wightman function
(\ref{ccr21}).

Turn now to the above mentioned Wick
rotation of Green's functions of Euclidean quantum fields to
causal forms on the Minkowski space.

Since the Minkowski space $X$ and its Euclidean partner $Y$ in
$\Bbb C^4$ have the same spatial subspace, we
further omit the dependence on  spatial coordinates.
Therefore, let us consider the complex plane $\Bbb C^1=X\oplus iZ$ of the time
$x$ and the Euclidean time $z$ and the complex plane $\Bbb C_1=P\oplus iQ$
of the associated momentum coordinates $p$ and $q$.

Let $W(q)\in S'(Q)$ be a tempered distribution such that
\mar{036}\beq
W=W_++W_-,\qquad W_+\in S'(\ol Q_+), \qquad W_-\in S'(\ol Q_-).\label{036}
\eeq
For instance, $W(q)$ is an ordinary function at $0$.
For every test function $\psi_+\in S(X_+)$, we have
\mar{037}\ben
&&\frac{1}{2\pi}\op\int_{\ol Q_+}W(q)\wh\psi_+(q)dq
=\frac{1}{2\pi}\op\int_{\ol
Q_+}dq\op\int_{X_+}dx [W(q)\exp(-qx)\psi_+(x)]=\nonumber\\
&&\qquad\frac{1}{(2\pi)^2}\op\int_{\ol Q_+}dq\op\int_Pdp\op
\int_{X_+}dx[W(q)\psi^F_+(p)\exp(-ipx-qx)]=\nonumber\\
&&\qquad \frac{-i}{(2\pi)^2}\op\int_{\ol Q_+}dq
\op\int_Pdp[W(q)\frac{\psi^F_+(p)}{p-iq}]=\frac{1}{2\pi}
\op\int_{\ol Q_+}W(q)\psi^L_+(iq)dq, \label{037}
\een
due to the fact that the FL transform
$\psi^{FL}_+(p+iq)$ of the function $\psi_+\in S(X_+)\subset S'(X_+)$
exists and that it is holomorphic on the tube $P+iQ_+, Q_+\subset
Q_{\phi_+}$. Moreover, $\psi^{FL}_+(p+i0)=\phi^F_+(k)$, and the function
$\wh\psi_+(q)=\psi^{FL}_+(-q)$ can be regarded as the Wick rotation
of the test function
$\psi_+(x)$. The equality (\ref{037}) can be brought into the form
\mar{038}\ben
&&\frac{1}{2\pi}\op\int_{\ol Q_+}W(q)\wh\psi_+(q)dq=
\op\int_{X_+}\wh W_+(x)\psi_+(x)dx, \label{038}\\
&&\wh W_+(x)=\frac{1}{2\pi}\op\int_{\ol Q_+}\exp(-qx)W(q)dq, \qquad
x\in X_+. \nonumber
\een
It associates to a distribution $W(q)\in S'(Q)$ the distribution
$\wh W_+(x)\in S'(X_+)$, continuous with respect to
the coarsen topology on $S(X_+)$.

For every test function $\psi_-\in S(X_-)$, the similar relations
\mar{039}\ben
&&\frac{1}{2\pi}\op\int_{\ol Q_-}W(q)\wh\phi_-(q)dq=
\op\int_{X_-}\wh W_-(x)\phi_-(x)dx, \label{039}\\
&&\wh W_-(x)=\frac{1}{2\pi}\op\int_{\ol Q_-}\exp(-qx)W(q)dq, \qquad 
x\in X_-, \nonumber
\een
hold. Combining (\ref{038}) and (\ref{039}), we obtain
\mar{040}\ben
&&\frac{1}{2\pi}\op\int_QW(q)\wh\psi(q)dq=\op\int_X\wh W(x)\psi(x)dx, 
\label{040}\\
&&\wh\psi=\wh\psi_++\wh\psi_-, \qquad \psi=\psi_++\psi_-,  \nonumber
\een
where $\wh W(x)$ is a linear functional on functions $\psi\in S(X)$, which
together with all derivatives vanish
at $x=0$. One can think of
$\wh W(x)$ as being the Wick rotation of the
distribution (\ref{036}). One should additionally define $\wh W$ at the point
$x=0$ in order to make it to a functional on the whole space $S(X)$. This
is the well-known ambiguity of chronological forms in quantum field theory.

\end{document}